\documentstyle[12pt,psfig]{article}
\newcommand{\bino}{\widetilde{B}}
\newcommand{\gsim}{ \mathop{}_{\textstyle \sim}^{\textstyle >} }
\newcommand{\lsim}{ \mathop{}_{\textstyle \sim}^{\textstyle <} }
\newcommand{\vev}[1]{ \left\langle {#1} \right\rangle }
\newcommand{\nn}{\nonumber}

\newcommand{\barr}[1]{ \overline{{#1}} }
\newcommand{\tb}{\tan \beta}
\newcommand{\stau}{\widetilde{\tau}}
\newcommand{\staul}{\widetilde{\tau}_{1}}
\newcommand{\sel}{\widetilde{e}_{1}}
\newcommand{\smul}{\widetilde{\mu}_{1}}
\newcommand{\ra}{\rightarrow}

\begin{document}
\baselineskip 0.6cm
\renewcommand{\thefootnote}{\fnsymbol{footnote}}
\def\tr{\mathop{\rm tr}\nolimits}
\def\Tr{\mathop{\rm Tr}\nolimits}
\def\Re{\mathop{\rm Re}\nolimits}
\def\Im{\mathop{\rm Im}\nolimits}
\setcounter{footnote}{1}

\begin{titlepage}

\begin{flushright}
UT-892\\
\end{flushright}
 
\vskip 2cm
\begin{center}
{\large \bf Cosmological Gravitino Problem \\
in Gauge-Mediated Supersymmetry Breaking Models}
\vskip 1.2cm
T. Asaka, K. Hamaguchi, and Koshiro Suzuki

\vskip 0.4cm

{\it Department of Physics, University of Tokyo, \\
     Tokyo 113-0033, Japan}\\
\vskip 0.2cm
(May 15, 2000)
\vskip 2cm
\abstract{ We investigate the cosmological gravitino problem in
gauge-mediated supersymmetry breaking models, where the gravitino
becomes in general the lightest supersymmetric particle (LSP).  In
order to avoid the overclosure of the stable gravitino, the reheating
temperature of inflation $T_R$ should be low enough.  Furthermore, if
the gravitino mass is larger than about 100 MeV, the decay of the
next-to-LSP (NLSP) into the gravitino may modify disastrously the
abundances of the light elements predicted by the big-bang
nucleosynthesis (BBN).  We consider the case in which the lighter stau
is the NLSP and derive cosmological constraints from the BBN on the
stau NLSP decay. We obtain a lower bound on the mass of stau
$m_{\staul}$, which is more stringent than the current experimental
limit $m_{\staul} > 90$ GeV for the gravitino mass region $m_{3/2}
\gsim 5$ GeV. This lower bound, together with the overclosure
constraint on the stable gravitino, gives an upper bound on $T_{R}$.
We find that the reheating temperature can be as high as
$10^9$--$10^{10}$ GeV for $m_{3/2} \simeq 5$--100 GeV.  }
\end{center}
\end{titlepage}

\renewcommand{\thefootnote}{\arabic{footnote}}
\setcounter{footnote}{0}

\section{Introduction}

~~~~ Supersymmetry (SUSY) has been considered as an attractive
candidate for physics beyond the standard model.  Many mechanisms have
been proposed for SUSY breaking and its transmission to our sector.
Among them, the gauge-mediated SUSY breaking (GMSB) models
\cite{dns,GR} are fascinating since they beautifully solve the problem
of flavor changing processes inherent in the SUSY standard model.
Moreover, GMSB models are determined by a few parameters, and thus
have high predictability. In general, they predict that the gravitino
is the lightest SUSY particle (LSP) and the next-to-lightest SUSY
particle (NLSP) is either the lightest neutralino (almost purely bino
$\bino$) or the lighter stau $\staul$ \cite{dimopoulos_thomas_wells}.
  This mass spectrum gives
striking features to the low energy phenomenology.  Here we focus on
the cosmological aspects of GMSB models.

The light stable gravitino
\footnote{ We assume here that the $R$-parity is exact and hence the
LSP gravitino is stable.}
gives rise to a serious cosmological
problem, known as the ``gravitino problem''
\cite{pp,moroi_murayama_yamaguchi,gouvea_moroi_murayama}. The energy
density of the gravitinos which are produced in the early universe may
exceed the present critical density if the gravitino mass is $m_{3/2}
\gsim$ 1 keV \cite{pp}.  Even if a primordial inflation is assumed,
this problem cannot be solved completely since gravitinos are
reproduced after the inflation ends.  At the reheating epoch,
gravitinos are produced by scatterings in the thermal bath. It has
been pointed out in Ref.~\cite{kallosh_kofman} that
gravitinos might also be produced non-thermally at the preheating
epoch. Although this non-thermal mechanism might dominate the thermal
production, we will not consider it, since it highly depends
on inflation models and we would like to have a general discussion.

Stable gravitinos which are produced thermally do not overclose the
universe if the reheating temperature of the inflation $T_{R}$ is low
enough. The upper bound on $T_{R}$ is given by
\cite{moroi_murayama_yamaguchi,gouvea_moroi_murayama}
\begin{equation}
 T_{R} \lsim 
\left\{
\begin{array}{ll}
 100 \;{\rm GeV} \;{\rm to}\; 1 \;{\rm TeV}  & {\rm for} \;\; 1 \;{\rm keV} \lsim m_{3/2} \lsim 100 \;{\rm keV} \\
 10^{8} \;{\rm GeV}\; \times 
 \left( \displaystyle{ \frac{m_{3/2}}{1 \;{\rm GeV}} } \right)
\left( \displaystyle{ \frac{m_{\bino}}{ 100 \;{\rm GeV}} } \right)^{-2}  
 & {\rm for} \;\; m_{3/2} \gsim 100 \;{\rm keV}
\end{array} ,
\right.
\label{UBTR}
\end{equation}
where $m_{\bino}$ denotes the bino mass.
Since this upper bound on $T_{R}$ is so severe for a lighter gravitino
mass region
\footnote{
Of course, there is no gravitino problem for $m_{3/2}$ $\lsim$ 1 keV
\cite{pp}.}
, inflation models which predict high reheating
temperatures are allowed only for a larger $m_{3/2}$ region.

However, when one considers a heavy gravitino (say, $m_{3/2}$
$\gsim$ 100 MeV) in GMSB, there is another cosmological
difficulty associated with the NLSP decay.
The NLSP decays into the LSP gravitino only through gravitational
interaction and its lifetime will be comparable to the big-bang
nucleosynthesis (BBN) era ($t \sim$1--$10^2$ sec).
The decay of the NLSP during or after the BBN is dangerous, since the
decay products might alter the abundances of the light elements, which
spoils the success of the BBN
\cite{BBNphoto,DEHS,reno_seckel}.
To avoid this difficulty, the lifetime
and the energy density of the NLSP are severely restricted
\cite{moroi_murayama_yamaguchi,bolz,holtmann,ghergetta_giudice_riotto}.

In this letter, we investigate cosmological difficulties associated
with the stable gravitino and derive the most conservative upper bound
on the reheating temperature.  Especially, we consider in detail
the cosmological consequences of the NLSP decay into the gravitino around
the BBN epoch.  Similar analysis had been made in
Ref.~\cite{ghergetta_giudice_riotto}, in which the BBN constraint is
obtained by considering mainly the effects of hadrons produced in the
NLSP decay.  However, the BBN with high energy hadron injection,
examined in Refs.~\cite{DEHS,reno_seckel}, has not completely
been settled yet, since there exist some uncertainties
\footnote{ For example, there exist uncertainties of the experimental
data of the hadron scattering processes, and also of the statistical
treatment of the errors.
}
.  Here, since we would like to obtain the most conservative
constraint on the reheating temperature, we will focus on the BBN
constraint coming from the photo-dissociation of the light
elements by the NLSP decay.  Furthermore, we consider the case when
the lighter stau $\staul$ is the NLSP.  This is because, compared
to the $\bino$ NLSP, the $\staul$ NLSP has larger annihilation
cross sections and so its abundance when it decays is smaller, which
results in a weaker constraint on the reheating temperature.
\section{Estimation of $\staul$ NLSP abundance}

~~~~ First of all, we discuss cosmological evolution of the $\staul$
NLSP and estimate its abundance when it decays, since the abundance is
crucial for obtaining BBN constraints.  Let us start from a brief
thermal history of $\staul$. By the time when the cosmic temperature
is comparable to the $\staul$ mass ($T \simeq m_{\staul}$), only the
$\staul$ NLSP among the SUSY particles is in thermal equilibrium with
the standard model particles. When $\staul$'s become non-relativistic
for $T$ $\lsim$ $m_{\staul}$, they pair-annihilate into standard model
particles and the abundance decreases exponentially. At $T \simeq
T_{f}$ ($T_{f}$: the freeze-out temperature
of $\staul$%
\footnote{ In our case, we find $T_f/m_{\staul} \simeq 1/28$--$1/25$
by numerical calculation.}
), $\staul$ decouples from the thermal bath and its abundance freezes
out. Finally, when the Hubble parameter $H$ becomes comparable to the
decay rate, $\staul$ decays into the gravitino.
The evolution of $\staul$ described above can be traced by solving the
Boltzmann equation.

Before writing down the equation, we must take care of the following two 
facts. 
The first one is that the masses of charged sleptons $\staul$,
$\smul$, and $\sel$ (the subscript 1 denotes the lighter mass
eigenstates) are almost degenerate in GMSB models.  This is because
they are mostly right-handed, and hence receive at the messenger scale
the same soft masses which are determined by the gauge quantum
numbers.  The mass differences between them at the weak scale are
induced from the renormalization group effects due to the leptonic
Yukawa couplings $y_{i}$ ($i= e,\mu,\tau$) and also from the
left-right mixings through $y_{i}$.  Considering the observed masses
of charged leptons, we safely neglect $y_{e}$ and $y_{\mu}$, and hence
the masses for $\smul$ and $\sel$, $m_{\smul}$ and $m_{\sel}$, are the
same.  On the other hand, the tau Yukawa coupling $y_{\tau}$ gives
negative contribution to the stau mass when it is evolved from the
messenger scale to the weak scale, and the left-right mixing between
the staus also decreases $m_{\staul}$. Both effects lead to the fact
that $m_{\staul}$ is always smaller than $m_{\smul (\sel)}$. This mass
splitting $\Delta m$ $\equiv$ $m_{\smul (\sel)} - m_{\staul}$ plays a
crucial role in calculating the $\staul$ abundance.  If $\Delta m$ is
very small, the relic abundance of $\staul$ is determined by not only
its annihilation processes, but also by the annihilation with $\smul$
and $\sel$, i.e., we should include the coannihilation effects
\cite{griest_seckel}.  We have verified numerically that these
coannihilation effects become significant for $\Delta m$ $\lsim$
$T_{f}$.  In GMSB models, both $\Delta m$
$\lsim$ $T_{f}$ and $\Delta m$ $\gsim$ $T_{f}$ are allowed%
\footnote{
The mass difference $\Delta m$ becomes larger as $\tb$ increases,
where $\tb \equiv v_u/v_d$ ($v_u$  and $v_d$ denote the vacuum 
expectation values of Higgs fields which couple to
up-type and down-type quarks, respectively).
}%
, so we respect both cases.

Secondly, if the bino mass $m_{\bino}$ is close to $m_{\staul}$, we
should also take into account the coannihilation of $\staul$ with
$\bino$, as well as with $\smul$ and $\sel$.  However, this effect
increases the abundance of $\staul$, which results in more stringent
upper bound on $T_{R}$ (see the following discussions).  Thus, we forbid
the coannihilation of $\staul$ with $\bino$ and restrict ourselves in
the parameter region
\begin{equation}
 m_{\bino} \gsim m_{\staul} + T_{f}.
\label{bino_lb}
\end{equation}
In the actual calculation below, we take the lower bound on $m_{\bino}$ 
as $m_{\bino} \geq 1.1 m_{\staul}$ to be conservative.

Now we are at the point to present the Boltzmann equation including the
coannihilation effects.
This equation describes the evolution of the total number density of
charged sleptons $n = n_{\staul}+n_{\staul^{*}}
+n_{\smul}+n_{\smul^{*}} +n_{\sel}+n_{\sel^{*}}$, where $n_{i}$ stands
for the number density of one slepton species $i$%
\footnote{
We distinguish particles from their anti-particles.}
\cite{griest_seckel},
\begin{equation}
 \frac{dn}{dt} = -3Hn - \vev{\sigma_{\rm eff} v }\left[ n^{2} - (n^{{\rm eq}})^{2}
\right],
\label{boltz3}
\end{equation}
where
\begin{equation}
 \vev{\sigma_{\rm eff}v} = \sum_{i,j}^{} \vev{ \sigma_{ij} v }
\frac{n_{i}^{{\rm eq}}}{n^{{\rm eq}}} \frac{n_{j}^{{\rm eq}}}{n^{{\rm eq}}}.
\label{sigma_eff}
\end{equation}
Here
$n^{\rm eq}$ ($n^{\rm eq}_{i}$) is the equilibrium value of $n$
($n_{i}$) and 
$v$ is the relative velocity of particles $i$
and $j$. The bracket denotes the thermal average and
$\sigma_{ij}$ is the total annihilation cross section of $i+j \ra X+X'$:
\begin{equation}
 \sigma_{ij} = \sum_{X,X'} \sigma( i+j \ra X+X'),
\end{equation}
where $X$ and $X'$ represent the possible standard model particles.
Note that the Boltzmann equation (\ref{boltz3}) also holds for the case where
the coannihilations of $\staul$ with $\smul$ and $\sel$ do not occur.

Next we discuss the thermal-averaged cross sections $\vev{\sigma_{\rm
eff} v}$ in Eq.~(\ref{boltz3}).  The thermal-averaged cross sections
$\vev{\sigma_{ij} v}$ in Eq.~(\ref{sigma_eff}) can be expanded in
terms of $T/m_{\staul}$ [see Eqs.~(\ref{staustau*}) -- (\ref{stausmu})
below].  Since the final abundance is determined by $\vev{\sigma_{\rm
eff} v}$ at $T \sim T_f$ and the freeze out temperature is typically
$T_{f} \simeq m_{\staul}/25$, it is sufficient only to consider the
leading term, i.e., the $s$-wave cross sections of the sleptons.
Note that the $s$-wave component of the thermal-averaged cross section
$\vev{\sigma_{ij} v}$ is equal to that of $\sigma_{ij}$.
The relevant (co)annihilation channels are
\begin{eqnarray}
&
({\rm I})&
\left\{
\begin{array}{l}
 \staul \staul^{*} \\
 \smul \smul^{*} \\
 \sel \sel^{*}
\end{array}
\right.
 \ra \gamma \gamma, Z \gamma, ZZ, W^{+}W^{-}, f \barr{f}, h^{0} h^{0}
 \nn \\
&
({\rm II})&
\left\{
\begin{array}{l}
\staul \staul \ra \tau~ \tau  \\
\staul^{*} \staul^{*} \ra \barr{\tau}~ \barr{\tau} \\
\smul \smul  \ra \mu~ \mu  \\
\smul^{*} \smul^{*} \ra \barr{\mu}~ \barr{\mu}  \\
\sel \sel \ra e~ e  \\
\sel^{*} \sel^{*} \ra \barr{e}~ \barr{e}  \\
\end{array}
\right.
 \nn \\
&
({\rm III})&
\left\{
\begin{array}{l}
\staul \smul \ra \tau~ \mu  \\
\staul^{*} \smul^{*} \ra \barr{\tau}~ \barr{\mu}  \\
\staul \sel \ra \tau~ e  \\
\staul^{*} \sel^{*} \ra \barr{\tau}~ \barr{e}  \\
\smul \sel \ra \mu~ e  \\
\smul^{*} \sel^{*} \ra \barr{\mu}~ \barr{e}  \\
\end{array}
\right.
\nn \\
&
({\rm IV})&
\left\{
\begin{array}{l}
\staul \smul^{*} \ra \tau~ \barr{\mu}, \; \nu_{\tau}~ \barr{\nu}_{\mu} \nn \\
\staul^{*} \smul \ra \barr{\tau}~ \mu, \; \barr{\nu}_{\tau}~ \nu_{\mu} \nn  \\
\staul \sel^{*}  \ra \tau~ \barr{e}, \; \nu_{\tau}~ \barr{\nu}_{e} \nn \\
\staul^{*} \sel  \ra \barr{\tau}~ e, \; \barr{\nu}_{\tau}~ \nu_{e} \nn \\
\smul \sel^{*}  \ra \mu~ \barr{e}, \; \nu_{\mu}~ \barr{\nu}_{e} \nn \\
\smul^{*} \sel  \ra \barr{\mu}~ e, \; \barr{\nu}_{\mu}~ \nu_{e} \nn
\label{decaymode}
\end{array}
\right.
\nn 
\end{eqnarray}
Here $f$ denotes the ordinary quarks and leptons.

\begin{figure}[t]
 \centerline{\psfig{figure=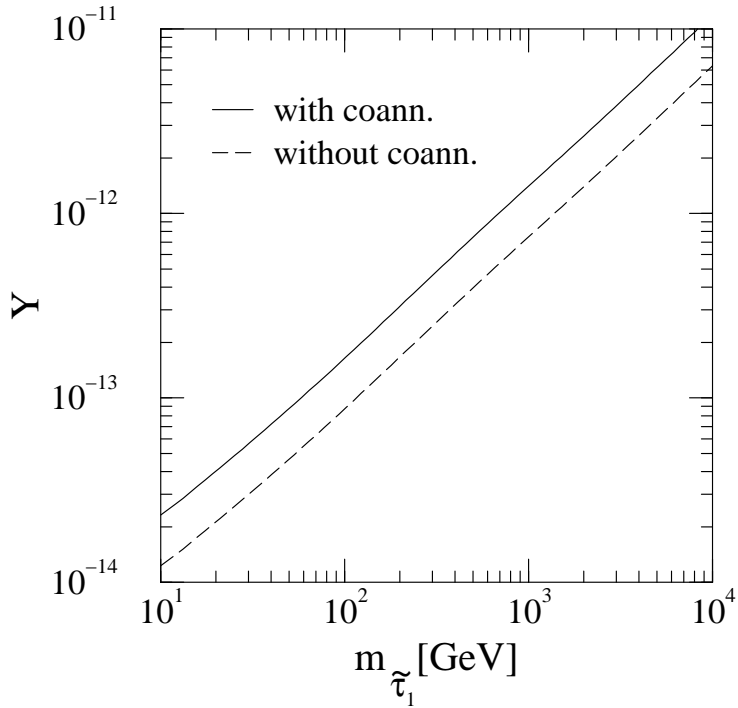,height=9.5cm}}
\caption{
The final $\staul$ abundance $Y$.
The solid line represents the abundance for the case with
 coannihilation  and the dashed line represents that for the case
 without coannihilation.}
\label{fig.Ystau}  
\end{figure}

We calculate cross sections for all the processes listed in
(I)-(IV). The processes (I) and (II) represent the
annihilation of each slepton species, while (III) and (IV) 
describe coannihilation processes. In the following we briefly explain
their features. First of all, let us estimate the cross
sections for processes (I). Although they have many final states, it
turns out that most of them give very small contributions to the total
cross section.
Processes which are induced by the tau Yukawa coupling and/or the
left-right mixing of sleptons are
suppressed due to their smallness%
\footnote{The mixing angles $\alpha_{i}$ of sleptons are in
general small in GMSB models ($\sin \alpha_{i} \ll 1$).
However, in the extremely large $\tan \beta$ region,
the stau mixing angle $\alpha_{\stau}$ can be large as $\sin \alpha_{\stau}
\simeq {\cal O}$(0.1).
Although
$\vev{\sigma_{\staul \staul^{*}} v }/ \sigma (\staul \staul^{*} \ra
\gamma \gamma)$ might become of ${\cal O}$(10) as opposed to
Eq.~(\ref{staustau*}) in this case,
we find that the final upper bound on $T_{R}$ only increases 
by less than a factor of 2.}%
.
Furthermore, in calculating $s$-wave cross sections, derivative
couplings of the sleptons vanish.  From these considerations, we find
that $\gamma \gamma$ and $Z \gamma$ are the dominant channels%
\footnote{ The ratios of the cross sections for the dominant processes
$\gamma \gamma$,~$Z \gamma$, and $ZZ$ are 1.0:0.6:(0.1--0.2).
The contributions of the other channels
are less than a few percents.}%
, and $\langle \sigma_{\staul \staul^{*}} v \rangle$ is approximately
given by%
\footnote{
Actually, if $m_{\staul} < m_{Z}/2$, the $Z \gamma$ channel is forbidden
kinematically.
Furthermore, if $m_{\staul} \simeq m_{h^{0}}/2$ ($m_{h^{0}}$: mass of
the lightest Higgs boson $h^{0}$), the annihilation cross section
is enhanced by the pole contribution of $h^{0}$.  These effects may
change the value of $Y$ in Fig.~\ref{fig.Ystau}.
However, we simply assume Eq.~(\ref{staustau*}), since the
mass region of $m_{\staul} < $ 90 GeV is excluded from the current
experimental limit.}%
:
\begin{equation}
 \vev{\sigma_{\staul \staul^{*}} v } \simeq 2 \sigma(\staul \staul^{*} \ra \gamma \gamma) = \frac{4\pi \alpha_{\rm em}^{2}}{m_{\staul}^{2}} + \; {\cal O}\left( \frac{T}{m_{\staul}} \right).
\label{staustau*}
\end{equation}
Next, we turn to the processes (II), which have only
$t$-channel and $u$-channel diagrams.
Diagrams induced by the exchange of Higgsino and neutral Wino
can be safely neglected due to the smallness of $y_{\tau}$ and
left-right mixing, respectively. Then, there is only $\bino$
contribution and its cross section is calculated as
\begin{eqnarray}
\vev{\sigma_{\staul \staul} v} &\simeq&
 \frac{16\pi \alpha_{\rm em}^{2} m_{\bino}^{2}}{\cos^{4}\theta_{W} 
 ( m_{\staul}^{2} + m_{\bino}^{2})^{2} } 
  + \; {\cal O}\left( \frac{T}{m_{\staul}} \right).
\label{staustau}
\end{eqnarray}
The annihilation channels listed in (III) can be calculated in the
same way as
\begin{eqnarray}
 \vev{\sigma_{\staul \smul} v} &\simeq& 
 \frac{8 \pi \alpha_{\rm em}^{2} m_{\bino}^{2} }
      { \cos^{4}\theta_{W}
       ( m_{\staul}^2 + m_{\bino}^{2} )^{2} }
 + \; {\cal O}\left( \frac{T}{m_{\staul}} \right),
\label{stausmu}
\end{eqnarray}
for $m_{\smul} = m_{\staul}$.
Cross sections for the processes (IV) are also dominated by the $\bino$
exchange. However, the left-right mixings
of sleptons are necessary for these processes, and so they give only
tiny contributions to the total thermal-averaged cross section
$\vev{\sigma_{\rm eff} v}$.
Cross sections for other processes in (I)--(IV) are calculated
similarly.
From the above arguments, we conclude that all the annihilation
cross sections are almost determined by only the masses of the bino
and the sleptons.
 
Finally, including all the cross sections described above, we solve the
Boltzmann equation (\ref{boltz3}) numerically.
Here we take $m_{\bino} = 1.1 ~m_{\staul}$ [see Eq.~(\ref{bino_lb})].
In this case the (co)annihilation cross sections take their maximal
values, which gives the smallest $\staul$ abundance.
In Fig.~\ref{fig.Ystau} the result is shown in terms of $Y$, which is
defined by $Y \equiv n/s = \sum_{i} n_{i}/s$ with the total
entropy density of the universe $s$.
Here are some comments. We calculate both of the cases with and
without coannihilation effects as mentioned before.  If the mass
difference between $\smul (\sel)$ and $\staul$ is large enough
($\Delta m \gsim T_{f}$), $\smul$ and $\sel$ decay into $\staul$ and
disappear from the thermal bath before $\staul$ freezes out.
Therefore, the coannihilation processes become ineffective.  In this
case, the abundance becomes $Y \simeq (n_{\staul} +n_{\staul^{*}})/s$
for $T \lsim T_{f}$, since $n_{\smul} \simeq n_{\sel} \simeq 0$. For
$T \lsim T_f$ the number density of $\staul$ and the entropy density
decrease at the same rate as the universe expands, and $Y$ takes a
constant value until $\staul$ decays.  In fact, Fig.~\ref{fig.Ystau}
shows the value of $Y$ just before the decay of $\staul$.  On the
other hand, when $\Delta m \lsim T_{f}$, $\smul$ and $\sel$ are still
in thermal equilibrium at $T \simeq T_{f}$ due to the effect of
inverse decays, and thus we should consider the coannihilations.  Note
that $Y$ takes a constant value after sleptons decouple from the thermal
bath even in this case. This is because $Y$ is an invariant parameter
against the expansion of the universe and also because $n= n_{\staul}
+n_{\staul^{*}} +n_{\smul}+ n_{\smul^{*}}+ n_{\sel}+ n_{\sel^{*}}$
does not change by the decays of $\smul$ and $\sel$.  Therefore, the
final abundance for $\staul$ is given by $Y$ for both cases
with and without coannihilation effects.

As shown in Fig.~\ref{fig.Ystau}, the final abundance with coannihilation
is larger than that without coannihilation. This can be understood as
follows: Imagine when the coannihilation cross sections in (III) and
(IV) are extremely large. In this case, the final abundance $Y$ would be
smaller than that without coannihilation. On the other hand, if the
coannihilation cross sections are zero, $Y$ increase by a factor of 3
because the relevant degrees of freedom is now 6, not 2.  In our case,
coannihilation cross sections in (III) and (IV) are smaller than those
in (I) and (II), and the final abundance with coannihilation is about
twice as large as that without coannihilation.

We present in Fig.~\ref{fig.Ystau} the results for the two extreme cases
$\Delta m \simeq 0$ and $\Delta m \gg T_f$, for illustration and for
comparison. In fact, we find that the final abundance of stau for the case $0 \lsim
\Delta m \lsim T_{f}$ falls between the two lines in Fig.~\ref{fig.Ystau}.

Before closing this section, we should comment on the case in which the
mass difference between $\staul$ and $\smul (\sel)$ is extremely small.
If the mass difference is smaller than the tau mass (of course we should
include the coannihilation effects in this case), the decay channel
$\smul (\sel) \ra \staul \barr{\tau} \mu (e)$ is kinematically
forbidden.  Then, $\smul$ and $\sel$ may dominantly decay into
gravitinos%
\footnote{ The left-right mixing allows $\smul$ to decay through $\smul
\ra \nu_{\mu} \staul \barr{\nu}_{\tau}$. If this is the main decay
channel, the following discussion does not change, since
$\smul$ has already decayed into $\staul$ before the $\staul$
decays.}
and so we should consider the effect of the $\smul(\sel)$ decay on the
BBN, as well as $\staul$.  However, we simply neglect this
possibility, since the final results do not change much.
\section{BBN constraint on $\staul$ NLSP decay}

~~~~
We are now ready to investigate the cosmological consequence of 
the decay of the $\staul$ NLSP into the gravitino at the BBN epoch.
The $\staul$ decay rate  is estimated, for $m_{\staul} \gg m_{3/2}$, as
\begin{equation}
 \Gamma_{\staul}
\simeq \frac{1}{48\pi}\frac{m_{\staul}^{5}}{m_{3/2}^{2} M_{*}^{2}},
\label{stau-decaywidth}
\end{equation}
with $M_\ast = 2.4 \times 10^{18}$ GeV, 
and the lifetime is given by
\begin{eqnarray}
    \tau_{\staul} \simeq
    6 \times 10^{4} ~\mbox{sec}
    \left( \frac{m_{3/2}}{1~\mbox{GeV}} \right)^2
    \left( \frac{m_{\staul}}{100~\mbox{GeV}} \right)^{-5}.
\label{stau_lifetime}
\end{eqnarray}
Therefore, $\staul$ is found to be a long-lived particle
and to decay during or even after the BBN epoch when $m_{3/2} \gsim$ 10
MeV for $m_{\staul}$ = 100 GeV, and hence we
should seriously consider the effects of its decay on the BBN.
%
\begin{figure}[t]
 \centerline{\psfig{figure=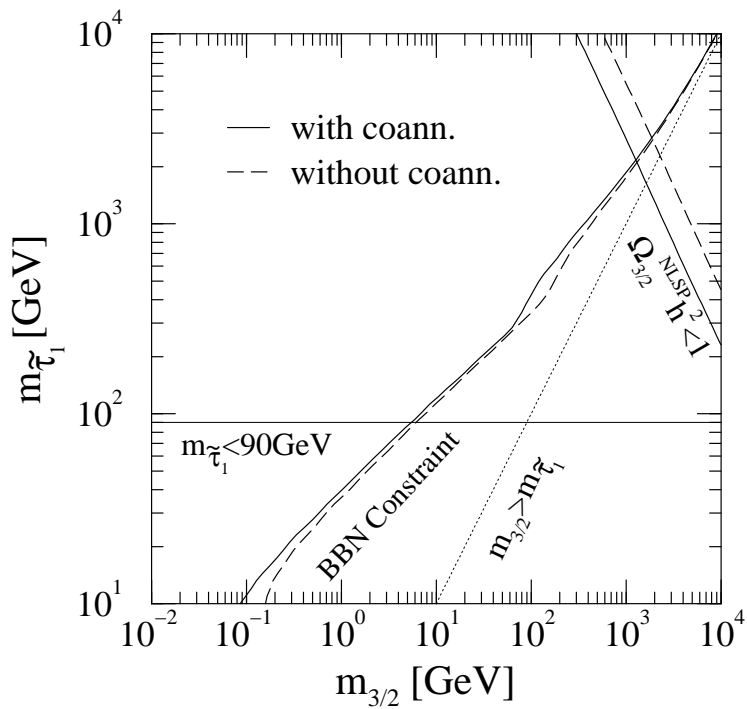,height=9.5cm}}

\caption{ Various bounds on the $\staul$ mass.  The
lower bounds on $m_{\staul}$ from the BBN photo-dissociation effect
are shown by the thick solid and dashed lines for the cases with and
without the coannihilation effects, respectively.  The experimental 
lower bound on $m_{\staul}$ is shown by thin solid line.
We also present the upper bounds on $m_{\staul}$ from the constraint
$\Omega_{3/2}^{\rm NLSP} h^2 < 1$ by the thick solid
and dashed lines for the cases with and without the coannihilation
effects, respectively.  }
\label{fig.mstau_mgrav}
\end{figure}
%

The energetic $\tau$'s produced by the $\staul$ decay trigger the
electro-magnetic (EM) cascade processes and induce high energy photons.
These photons may be abundant enough to destroy or overproduce various
light elements (D, $^3$He, $^4$He, etc.)  synthesized by the BBN.  In
order to keep the success of the BBN, the energy density (per the
entropy density) of the extra EM particles emitted after the BBN era
(i.e., $t \gsim 10^4$ sec) is severely constrained \cite{BBNphoto}.
Here it should be noted that not all the energy of tau contribute to
these photo-dissociation processes.  This is because the tau decays
before it causes the EM cascade processes and the produced neutrinos
give no effects on the photo-dissociation processes~\footnote{In fact,
the high energy neutrinos produced by the tau decay also causes the EM
cascade processes by scattering with the background neutrinos. However,
this effect is less significant \cite{neu_dis}.}. By using the
result of the recent analysis in Ref.~\cite{holtmann} (see Fig.16 in
Ref.~\cite{holtmann}), we can obtain an upper bound on
\begin{eqnarray}
 \frac{\rho_{\staul}}{s} = m_{\staul} Y,
\end{eqnarray}
for a given lifetime of $\staul$. Here $\rho_{\staul}$ denotes the
energy density of $\staul$.  When one fixes the gravitino mass, this
bound is translated into the lower bound on $m_{\staul}$, since both
the abundance and the lifetime of $\staul$ are determined by its mass
$m_{\staul}$. The obtained result is found in
Fig.~\ref{fig.mstau_mgrav}.

You can see that this BBN photo-dissociation constraint gives a more
stringent lower bound on $m_{\staul}$ for $m_{3/2} \gsim 5$ GeV,
compared to  the experimental bound 
$m_{\staul} > 90$ GeV.
This result will help us to estimate the upper bound on the reheating
temperature in the next section.

\section{Gravitino problem and constraint on $T_{R}$}

~~~~
Finally, we discuss the cosmological gravitino problem
and obtain an upper bound on the reheating temperature of inflation.
At the reheating epoch after the inflation ends, 
gravitinos are produced thermally by scatterings with particles
in the hot plasma of the universe%
\footnote{
As mentioned in Sec.~1, we discard the non-thermal
production of gravitinos at the preheating epoch in order to obtain the 
most conservative result.}%
.
The relic abundance of the gravitino is given by
\cite{moroi_murayama_yamaguchi,gouvea_moroi_murayama}
\begin{equation}
 \label{om-th}
 \Omega_{3/2}^{\rm th} h^2 \simeq 
 0.3 \left( \frac{m_{3/2}}{1~\mbox{GeV}} \right)^{-1}
  \left( \frac{m_{\bino}}{100~\mbox{GeV}} \right)^2
  \left( \frac{T_R}{10^8~\mbox{GeV}} \right),
\label{grav_th}
\end{equation}
where $h$ is the present Hubble parameter in unit of 100km/sec/Mpc, and
$\Omega_{3/2}^{\rm th} = \rho_{3/2}^{\rm th}/\rho_{c}$
($\rho_{3/2}^{{\rm th}}$ is the present energy density of the thermally
produced gravitinos and $\rho_{c}$ is the critical density of the
present universe).  It is found from Eq.~(\ref{om-th}) that the
overclosure limit of $\Omega_{3/2}^{\rm th} < 1$ puts an upper bound on
$T_R$ as shown in Eq.~(\ref{UBTR}).  Notice that this upper bound on
$T_R$, if one fixes the $\bino$ mass, becomes more severe for the
lighter gravitino mass region, and the highest reheating temperature
allowed in GMSB models is achieved for a relatively heavy gravitino
mass.

Furthermore, it should be noted that gravitinos are also 
produced by the $\staul$ NLSP decays.
Because one gravitino is produced per a $\staul$ decay, we find that
\begin{equation}
 \Omega_{3/2}^{\rm NLSP} 
 \equiv
 \frac{ \rho_{3/2}^{ \rm NLSP} }{ \rho_c }
 =
  \frac{m_{3/2} Y }{\rho_c / s_0},
\label{grav_nlsp}
\end{equation}
where $\rho_{3/2}^{\rm NLSP}$ is the present energy density of 
the gravitinos produced by the $\staul$ decays,
and $s_0$ denotes the entropy density of the present universe.  In
Fig.~\ref{fig.mstau_mgrav} we also show the upper bound on
$m_{\staul}$ from the constraint $\Omega_{3/2}^{\rm NLSP} h^2< 1$.  It
is found that the gravitino mass is bounded from above as
$m_{3/2}\lsim 1$ TeV when combined with the lower bound on
$m_{\staul}$ from the BBN photo-dissociation effects.

%
\begin{figure}[t]
 \centerline{\psfig{figure=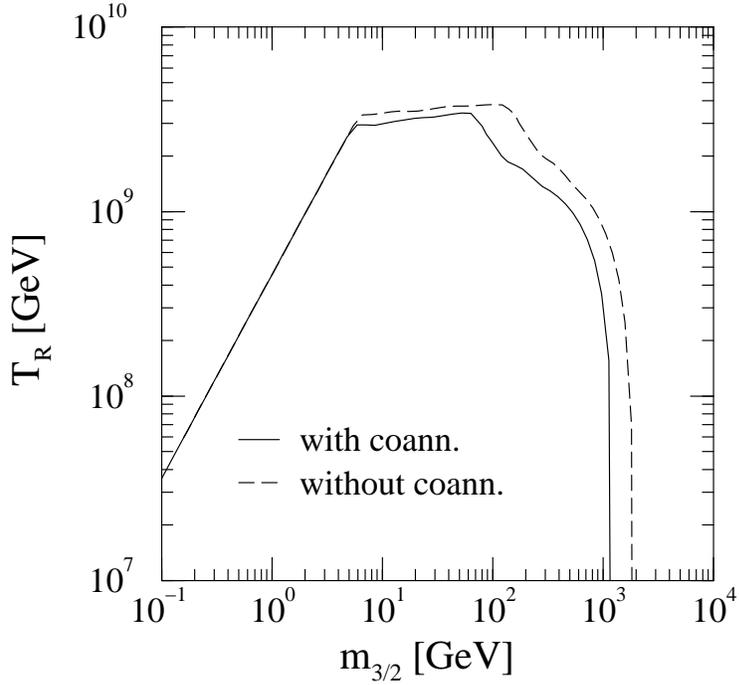,height=9.5cm}}
\caption{ The upper bounds on $T_{R}$.  The bounds are shown by the
thick solid and dashed lines for the cases with and without the
coannihilation effects, respectively.}  \label{fig.TR}
\end{figure}
%

Therefore, the total relic abundance of gravitinos in GMSB models is
now given by $\Omega_{3/2}^{\rm tot}= \Omega_{3/2}^{\rm th} +
\Omega_{3/2}^{\rm NLSP}$, as long as the decay of the $\staul$ NLSP
takes place within the age of the universe.  From the lower bound on
the $\staul$ mass obtained in the previous section, we can estimate
the most conservative upper bound on the reheating temperature in the
following way.  
Since the abundance $\Omega_{3/2}^{\rm th}$ 
depends on the bino mass $m_{\bino}$ as shown in
Eq.~(\ref{grav_th}), the weakest bound on $T_R$ is
obtained by the lowest value of $m_{\bino}$.
On the other hand, the lower bound on the mass
of the $\staul$ NLSP represented in Fig.~\ref{fig.mstau_mgrav} is
nothing but the lower bound on $m_{\bino}$.  
Now we are working in the
parameter region of $m_{\bino}$ in Eq.~(\ref{bino_lb}) in order to
forbid the coannihilation of $\staul$ with $\bino$.  Therefore, the
lowest value of $m_{\bino}$ coming from the $m_{\staul}$ constraint
gives the most conservative upper bound on the
reheating temperature
\footnote{ In fact, such a value of $m_{\bino}$ makes
the $\staul$ abundance smallest.
(See discussions in Sec.~2.)
}
.  
The result is found in Fig.~\ref{fig.TR},
where we take $h=1$ for simplicity
\footnote{
The upper bound on $T_R$ becomes more stringent for a smaller $h$.
}
.

Note that the gravitino mass region of $m_{3/2} \gsim 1$ TeV is
excluded, since the gravitinos produced by the decay of $\staul$
overclose the present universe ($\Omega_{3/2}^{\rm NLSP} > 1$).  On the
other hand, $\Omega_{3/2}^{{\rm NLSP}}$ plays no role in putting an
upper bound on $T_{R}$ for a lighter gravitino mass region of $m_{3/2}
\lsim 500$ GeV.

You can see that the upper bound on $T_R$ is almost proportional to
$m_{3/2}$ for $m_{3/2}$ $\lsim$ 5 GeV.  In this gravitino mass region,
the experimental bound on $m_{\staul}$ gives the lowest value of
$m_{\bino}$ (see Fig.~\ref{fig.mstau_mgrav}).  The upper bound on
$T_R$, thus, is the conventional one Eq.(\ref{UBTR}) with $m_{\bino}
\simeq 100$ GeV.  However, when $m_{3/2}$ $\simeq$ 5 GeV--100 GeV, the
upper bound on $T_R$ takes an almost constant value of
$10^9$--$10^{10}$ GeV.  In this region the bino mass which gives the
highest value of $T_R$ is determined by the lower bound on
$m_{\staul}$ from the BBN photo-dissociation constraint.  Since this
lower bound on $m_{\staul}$ is more stringent than the experimental
limit, the upper bound on $T_R$ also becomes more stringent than the
conventional one. From Fig.2, the lower  bound on the stau mass,
i.e., the lower bound on the bino mass is almost proportional to
$(m_{3/2})^{1/2}$ for $m_{3/2} =$ 5--100 GeV. Thus, it is found from
Eq.(12) that the upper bound on $T_{R}$ becomes almost
constant. Therefore, in GMSB models, the reheating
temperature can be taken as high as $T_R \simeq 10^9$--$10^{10}$ GeV
for $m_{3/2} \simeq 5$--100 GeV.

\section{Conclusions and discussions}

~~~~ In this letter, we have considered the cosmological gravitino
problem in GMSB models with the $\staul$ NLSP.  Especially, we have
investigated in detail the cosmological consequence of the $\staul$
decay soon after the BBN epoch.  Since the $\staul$ abundance when it
decays is crucial for this discussion, we have solved numerically the
Boltzmann equations for both cases with and without coannihilation
effects, and obtained the final $\staul$ abundance.  We have found
that the BBN constraint on the $\staul$ abundance is translated into
the lower bound on $m_{\staul}$, and that the obtained bound is more
stringent for $m_{3/2} \gsim 5$ GeV than the current experimental
limit $m_{\staul} > 90$ GeV.  This gives some hints for GMSB models.
If $\staul$ was detected at future collider experiments, the observed
$\staul$ mass would enable us to set an upper bound on the gravitino
mass, and hence on the SUSY breaking scale.  It has also been found
that the gravitino mass region of $m_{3/2} \gsim$ 1 TeV is excluded
(although it might be marginal considering the SUSY flavor problem),
since the energy density of gravitinos produced by the decays of
$\staul$'s exceeds the present critical density.

By using the lower bound on $m_{\staul}$, we have obtained an upper
bound on the reheating temperature $T_R$ of inflation in order to
avoid the overclosure problem of the LSP gravitino produced in
thermal scatterings and also in the $\staul$ decay.  What we have
found is that the most conservative upper bound on $T_R$ in GMSB
models is $T_R \lsim 10^9$--$10^{10}$ GeV when $m_{3/2} \simeq 5$--100
GeV.  This upper bound on $T_R$ is weaker than those in the
conventional hidden sector SUSY breaking models to solve the
cosmological problem of unstable gravitinos ($T_R \lsim 10^6$ GeV and
$10^8$ GeV for $m_{3/2} \simeq$ 100--500 GeV and 500 GeV--1 TeV,
respectively \cite{holtmann}). Therefore, it helps us a lot to
build SUSY inflation models without the cosmological gravitino
problem.

Such a high reheating temperature is also promising for baryogenesis.
The relevant example is the leptogenesis \cite{fukugita_yanagida} via
decays of heavy Majorana neutrinos, which is very attractive from the
viewpoint of the observed tiny neutrino masses. The leptogenesis
requires a high reheating temperature to generate the observed baryon
asymmetry
of the universe
\footnote{ In the leptogenesis scenarios where heavy Majorana neutrinos
are thermally produced, the reheating temperature of $T_R \gsim 10^{10}$
GeV is required to induce the sufficient baryon
asymmetry\cite{buchmuler}. On the other hand, for the case when heavy
Majorana neutrinos are non-thermally produced in the inflaton decays,
the reheating temperature of $T_R \gsim 10^{6}$ GeV is
required~\cite{LG-infdecay}.}
.  Thus, the upper bound on the reheating temperature $T_R \lsim
10^9$--$10^{10}$ GeV obtained in this letter ensures some scenarios of
the leptogenesis to work without the gravitino problem.

The results we have found here are almost independent on models of the
GMSB, since the abundance of the $\staul$ NLSP is fixed by the
$\staul$ mass.  However, there are some loop-holes in which these
constraints can be avoided.  If one introduces the $R$-parity breaking,
the NLSP can decay before the BBN (i.e., $t \lsim$ 1 sec) and
evade the BBN constraints, and also the LSP gravitino can decay within
the age of the universe.  Furthermore, if one assumes the late-time
entropy production such as the thermal inflation \cite{Lyth-Stewart}
in the thermal history of the universe, the abundances of the NLSP and
also the LSP gravitino are diluted away so that we are free from these
constraints.

In the present analysis, we have considered only the photo-dissociation
effects of the decay of $\staul$ on the BBN to make a conservative
analysis.  As pointed out in Ref.~\cite{ghergetta_giudice_riotto} the
high energy hadrons produced by the $\staul$ decay might also be
dangerous.  Here we would like to briefly comment on their effects.  The
high energy hadrons, if they are produced at the time $t \sim 1$--$10^4$
sec, delay the freeze-out of the $p$-$n$ conversion and raise the
number ratio of $n$ to $p$, which leads to the overproduction of D and
$^4$He~\cite{reno_seckel}.  
Furthermore, if $\staul$ decays at the time $t \gsim 10^4$ sec,
the produced hadrons destroy the light elements synthesized by the BBN 
and modify their abundances (e.g.,
increases the $^7$Li abundance)~\cite{DEHS}.
These effects give the upper bound on the $\staul$
abundance.  
However, there are some uncertainties in the BBN with high
energy hadron injection.  
The relevant
hadron scattering cross sections, especially those of Li,
have not been observed experimentally in detail.
Furthermore, the statistical treatment for the estimation of the errors 
has not completely settled yet.  
Therefore, in the present analysis 
we do not include the BBN constraints from the hadron injection.

\section*{Acknowledgements}
We would like to thank T. Yanagida for various suggestions 
and stimulating discussions and also 
K. Kohri and Y. Nomura for useful comments.
This work was partially supported by the Japan Society for the
Promotion of Science (T.A. and K.H.).

%
\clearpage
%
%
%
\newcommand{\Journal}[4]{{\sl #1} {\bf #2} {(#3)} {#4}}
\newcommand{\PL}{\sl Phys. Lett.}
\newcommand{\PR}{\sl Phys. Rev.}
\newcommand{\PRL}{\sl Phys. Rev. Lett.}
\newcommand{\NP}{\sl Nucl. Phys.}
\newcommand{\ZP}{\sl Z. Phys.}
\newcommand{\PTP}{\sl Prog. Theor. Phys.}
\newcommand{\NC}{\sl Nuovo Cimento}
\newcommand{\MPL}{\sl Mod. Phys. Lett.}
\newcommand{\PRep}{\sl Phys. Rep.}


\begin{thebibliography}{99}
%
\bibitem{dns}
   M.~Dine, A.E.~Nelson, Y.~Shirman,
        {\PR} {\bf D51} (1995) 1362;
   M.~Dine, A.E.~Nelson, Y.~Nir, Y.~Shirman,
        {\PR} {\bf D53} (1996) 2658.
%
\bibitem{GR}
   For a review, see
   G.F.~Giudice, R.~Rattazzi,
        {\PRep} {\bf 322} (1999) 419.
%
\bibitem{dimopoulos_thomas_wells}
   S.~Dimopoulos, S.~Thomas and J.D.~Wells,
	{\NP} {\bf 488} (1997) 39.
%
\bibitem{pp}
   H.~Pagels, J.R.~Primack,
        {\PRL} {\bf 48} (1982) 223.
%
\bibitem{moroi_murayama_yamaguchi}
   T.~Moroi, H.~Murayama, M.~Yamaguchi,
        {\PL} {\bf B303} (1993) 289.
%
\bibitem{gouvea_moroi_murayama}
   A.~Gouv\^{e}a, T.~Moroi, H.~Murayama,
        {\PR} {\bf D56} (1997) 1281.
%
\bibitem{kallosh_kofman}
   R.~Kallosh, L.~Kofman, A.~Linde, A.V.~Proeyen,
        {\PR} {\bf D61} (2000) 103503;
   G.F.~Giudice, A.~Riotto, I.~Tkachev,
        {\it JHEP} {\bf 9908} (1999) 009.
%
\bibitem{BBNphoto}
   M.Y.~Khlopov, A.D.~Linde, 
        {\PL} {\bf B138} (1984) 265;
   J.~Ellis, J.E.~Kim, D.V.~Nanopoulos,
        {\PL} {\bf B145} (1984) 181;
   M.~Kawasaki, T.~Moroi,
        {\PTP} {\bf 93} (1995) 879.
%
\bibitem{DEHS}
   S.~Dimopoulos, R.~Esmailzadeh, L.J.~Hall, G.D.~Starkman,
        {\it Astrophys. J.} {\bf 330}, 545 (1988);
        {\NP} {\bf B311}, 699 (1989).
%

\bibitem{reno_seckel}
   M.H.~Reno, D.~Seckel,
        {\PR} {\bf D37} (1988) 3441.
%
\bibitem{bolz}
   M.~Bolz, W.~Buchm\"uller, M.~Pl\"umacher,
        {\PL} {\bf B443} (1998) 209.
%
\bibitem{holtmann}
   E.~Holtmann, M.~Kawasaki, K.~Kohri, T.~Moroi,
        {\PR} {\bf D60} (1999) 023506.
%
\bibitem{ghergetta_giudice_riotto}
   T.~Ghergetta, G.F.~Giudice, A.~Riotto,
        {\PL} {\bf B446} (1999) 28.
%
\bibitem{griest_seckel}
   K.~Griest, D.~Seckel,
        {\PR} {\bf D43} (1991) 3191.
%
\bibitem{neu_dis}
    J.~Gratsias, R.J.~Scherrer, D.N.~Spergel,
    {\PL} {\bf B262} (1991) 298;
    M.~Kawasaki, T.~Moroi,
    {\PL} {\bf B346} (1995) 27.
%
\bibitem{fukugita_yanagida}
   M.~Fukugita, T.~Yanagida,
        {\PL} {\bf B174} (1986) 45.
%
\bibitem{buchmuler}
   See, for recent reviews, 
   W.~Buchm{\" u}ler, M.~Pl{\"u}macher,
        {\PRep} {\bf 320} (1999) 329; 
   M.~Pl{\" u}macher,
        {\NP} {\bf B530} (1998) 207 and references there in.
%
\bibitem{LG-infdecay}
   K.~Kumekawa, T.~Moroi, T.~Yanagida,
        {\PTP} {\bf 92} (1994) 437;
   G.~Lazarides, 
        hep-ph/9904428 and reference therein;
   G.F.~Giudice, M.~Peloso, A.~Riotto, I.~Tkachev,
        {\it JHEP} {\bf 9908} (1999) 014;
   T.~Asaka, K.~Hamaguchi, M.~Kawasaki, T.~Yanagida,
        {\PL} {\bf B464} (1999) 12;
        {\PR} {\bf D61} (2000) 083512.
%
\bibitem{Lyth-Stewart}
   D.H.~Lyth, E.D.~Stewart,
        {\PRL} {\bf 75} (1995) 201;
        {\PR} {\bf D53} (1996) 1784.

%
\end{thebibliography}
\end{document}